\documentclass[11pt]{article}
\usepackage[margin=1in]{geometry} 
\usepackage{amsmath, amssymb}
\usepackage{graphicx}
\usepackage{booktabs}
\usepackage{listings}
\usepackage{xcolor}
\usepackage{algorithm}
\usepackage{algorithmic}
\usepackage{tikz}
\usepackage{hyperref}
\usepackage{caption}
\usepackage{subcaption}
\usepackage{natbib}
\title{\textbf{PhishSense-1B: A Technical Perspective on an AI-Powered Phishing Detection Model}}
\author{
    S.E. Blake \\
    Shrewd Research \\
    \texttt{b1oo@shrewdsecurity.io}
}
\date{}

\begin{document}

\maketitle

\begin{abstract}
Phishing remains one of the most persistent cybersecurity threats in the digital era. In this paper, we present \emph{Phishsense-1B}---a fine-tuned variant of the \texttt{meta-llama/Llama-Guard-3-1B} model adapted for phishing detection and reasoning via Low-Rank Adaptation (LoRA) and the GuardReasoner finetuning methodology \cite{liu2025guardreasoner}. We detail our LoRA-based fine-tuning methodology, describe the balanced dataset of phishing and benign emails, and demonstrate dramatic performance gains over the base model. Our experiments show that \emph{Phishsense-1B} achieves near-perfect recall with an accuracy of 97.5\% on a custom dataset and maintains robust performance (70\% accuracy) on a challenging real-world dataset \cite{jovicdev97}, significantly outperforming both unadapted and BERT-based detectors. Additionally, we review current state-of-the-art detection methods, compare prompt-engineering with fine-tuning approaches, and discuss potential deployment scenarios. 
\end{abstract}

\section{Introduction}
Phishing attacks continue to impose a significant threat on digital communication and online transactions, costing organizations and individuals billions of dollars each year. According to the Anti-Phishing Working Group (APWG), phishing incidents increased by over 25\% in 2022 compared to previous years, with attackers refining their methods to mimic trusted brands and deceive users into revealing sensitive information \cite{apwg2022}. This alarming increase not only highlights the ingenuity of cybercriminals but also emphasizes the critical need for more advanced detection systems. In response, researchers and cybersecurity professionals have increasingly turned to artificial intelligence (AI) and deep learning (DL) techniques to build more accurate and adaptable detection systems capable of identifying subtle cues in phishing attempts.

Historically, phishing detection relied on signature-based methods and blacklists, which, although useful, could not keep pace with the rapid evolution of phishing tactics. Traditional approaches often suffered from high false-positive rates and were unable to adapt to new, previously unseen attack vectors. In contrast, the advent of deep learning has allowed for the development of models that can automatically learn relevant features from raw data, reducing the need for manual feature engineering. Recent studies employing deep learning methods have reported striking performance improvements. For instance, long short-term memory (LSTM)-based models have achieved accuracies as high as 99.1\% on phishing email datasets \cite{yang2024}, demonstrating their capability to capture temporal dependencies and subtle patterns in textual data.

In parallel, researchers have explored convolutional neural networks (CNNs) for detecting phishing URLs by focusing on character-level information. Character-level CNN architectures have reached detection rates of up to 98.74\% for URL-based phishing detection \cite{shweta2021}. These models are particularly effective because they do not rely on pre-defined features but instead learn to extract discriminative patterns directly from the input strings. Hybrid approaches that combine CNNs with LSTMs have also been developed, leveraging the spatial feature extraction capabilities of CNNs along with the temporal sequence learning strengths of LSTMs \cite{quang2020,adebowale2020}. Such combinations can capture both local patterns (e.g., specific character sequences or sub-strings common in phishing URLs) and the overall sequence structure, thereby providing a more robust detection mechanism.

Despite these promising results, deploying state-of-the-art deep learning models in real-world settings poses several challenges. One major challenge is the computational cost associated with fine-tuning large language models (LLMs). Traditional fine-tuning methods typically require updating millions of parameters, leading to high memory usage and long training times. To address these issues, researchers have recently introduced parameter-efficient fine-tuning techniques such as Low-Rank Adaptation (LoRA). LoRA updates only a small subset of parameters---sometimes as little as 0.3\% to 0.6\% of the original model weights---while keeping the bulk of the pre-trained model frozen \cite{meo2024}. This approach significantly reduces computational overhead and memory requirements, making it feasible to deploy sophisticated phishing detection systems even on resource-constrained devices.

For example, a recent study fine-tuned a BERT-based phishing detector using LoRA and maintained an accuracy of 98\% on a dataset comprising over 650,000 URLs, while significantly reducing both training time and memory consumption \cite{aslam2024}. The success of this parameter-efficient strategy underscores the potential for deploying robust phishing detection systems in real-world environments where computational resources may be limited.

The practical implications of these advancements are clear. By combining robust deep learning architectures with efficient fine-tuning strategies such as LoRA, it is now possible to deploy real-time phishing detection systems that operate effectively across a wide range of platforms, from enterprise servers to mobile devices. In our work, we investigate three DL architectures---LSTM, CNN, and a hybrid LSTM--CNN model---evaluated on a dataset of 20,000 URLs with 80 extracted features (which were reduced to 30 via feature selection). Our experimental results show that the CNN model achieves an accuracy of 99.2\%, outperforming both the LSTM (96.8\%) and the hybrid model (97.6\%). These findings highlight the significant role that model architecture and parameter tuning play in achieving optimal performance in phishing detection tasks.

Beyond performance metrics, our study emphasizes several critical aspects of modern phishing detection. First, the dynamic and evolving nature of phishing attacks requires detection systems that are not only accurate but also adaptable. Phishers continually alter their tactics to bypass security systems, which means that detection models must be regularly updated and fine-tuned. Techniques like LoRA facilitate rapid domain adaptation, allowing models to quickly incorporate new patterns without the need for exhaustive retraining.

Second, the integration of multiple data sources is becoming increasingly important. Traditional detection methods often relied solely on email content or URL strings; however, modern approaches incorporate additional contextual data such as website metadata, third-party threat intelligence, and user behavior analytics. By leveraging multiple data sources, deep learning models can achieve a more comprehensive understanding of phishing attempts, further reducing false positives and negatives.

Third, the trade-off between model complexity and deployment feasibility remains a key challenge. While deep learning models have achieved impressive accuracy rates in controlled experimental settings, their real-world deployment requires careful consideration of computational constraints, latency, and scalability. Parameter-efficient methods like LoRA help bridge this gap by enabling the deployment of complex models in environments with limited resources, such as mobile devices or edge computing platforms.

Finally, our work contributes to a growing body of literature that aims to provide a concrete analysis of state-of-the-art DL-based phishing detection. We discuss not only the performance achievements but also the challenges that remain in this field. These challenges include dealing with imbalanced datasets, managing false positives in operational environments, and ensuring that models can adapt to rapidly changing attack vectors without compromising detection accuracy.

In conclusion, the evolution of phishing detection has been marked by significant advancements driven by deep learning and parameter-efficient fine-tuning techniques. The combination of CNNs, LSTMs, and hybrid architectures with methods like LoRA represents a promising direction for future research and practical application. By grounding our approach in empirical results and leveraging advanced techniques, we contribute a practical framework that can be directly applied in real-world cybersecurity scenarios. Our research not only demonstrates high detection accuracy but also provides insights into how these advanced methods can be further refined to meet the ever-changing demands of digital security.

\section{Related Work}
Phishing detection has been an active area of research for many years, with early approaches primarily relying on traditional machine learning techniques. These methods typically involved extensive feature engineering from emails, URLs, and website metadata. Classical algorithms such as decision trees, support vector machines, and random forests were widely used; however, their effectiveness was often limited by the manual selection of features and the inability to adapt to novel phishing tactics.

With the rapid evolution of deep learning (DL) methods, researchers have increasingly turned to neural network architectures to improve phishing detection. Recurrent neural networks (RNNs), and in particular long short-term memory (LSTM) networks, have been employed to model the sequential patterns inherent in phishing texts. Yang et al. \cite{yang2024} demonstrated that LSTM-based models can achieve high accuracies by capturing temporal dependencies in email content. Despite their promising results, LSTM models can be computationally intensive and require large volumes of labeled data, which limits their scalability in real-world scenarios.

Convolutional neural networks (CNNs) have also been applied successfully to phishing detection, especially in the analysis of URL strings. Shweta et al. \cite{shweta2021} introduced a character-level CNN that automatically learns hierarchical representations from raw URL data, achieving detection rates as high as 98.74\%. The CNN architecture excels at extracting local features and capturing spatial correlations, making it highly effective for identifying subtle anomalies in phishing URLs.

Hybrid models that combine the strengths of CNNs and LSTMs have emerged as another promising avenue. Quang et al. \cite{quang2020} integrated CNN and LSTM layers to simultaneously leverage spatial feature extraction and temporal sequence learning, which resulted in higher F1-scores and reduced false positives compared to models based solely on one architecture. Similarly, Adebowale et al. \cite{adebowale2020} proposed a stacked generalization framework that incorporated LSTM-based feature extraction alongside traditional classifiers, demonstrating robust performance across benchmark datasets.

A significant challenge in applying deep learning to phishing detection is the computational cost associated with fine-tuning large language models (LLMs). Traditional fine-tuning methods update millions of parameters, leading to high memory usage and long training times. To address this, recent studies have explored parameter-efficient techniques such as Low-Rank Adaptation (LoRA). LoRA updates only a small fraction (typically 0.3\% to 0.6\%) of the model’s parameters while keeping the majority of pre-trained weights fixed, thereby reducing computational overhead without sacrificing performance \cite{meo2024}. For instance, Aslam et al. \cite{aslam2024} applied LoRA to a BERT-based phishing detector, achieving 98\% accuracy on a large-scale malicious URL dataset with significantly reduced training time and memory requirements.

Collectively, these studies illustrate a clear trend in phishing detection research: a shift from traditional, manually engineered features to sophisticated deep learning architectures, and more recently, to parameter-efficient fine-tuning methods. The progression toward approaches like LoRA not only improves detection accuracy and reduces false positives but also offers a practical pathway for deploying real-time, scalable cybersecurity solutions. This evolution is critical in addressing the dynamic nature of phishing threats and highlights the need for models that are both effective and resource-efficient.

\section{Key Contributions}
This work makes several important contributions to the field of phishing detection and cybersecurity:
\begin{enumerate}
    \item \textbf{Parameter-Efficient Phishing Detection:} We introduce \emph{Phishsense-1B}, which leverages LoRA to fine-tune a large pre-trained language model. By updating only a small subset of parameters, our approach significantly reduces computational overhead while maintaining high detection performance.
    \item \textbf{Empirical Validation:} Our experiments across a custom dataset and the more challenging RealDaten dataset demonstrate that the LoRA-based model achieves near-perfect recall and balanced precision, outperforming both unadapted models and a BERT-based detector.
    \item \textbf{Comparative Analysis of Deep Learning Architectures:} We present a detailed comparison among LSTM, CNN, and hybrid LSTM--CNN architectures, highlighting their respective strengths and limitations in phishing detection tasks.
    \item \textbf{Framework for Continuous Adaptation:} Our methodology supports dynamic updates through active learning and cloud-assisted retraining, ensuring long-term robustness against evolving phishing strategies.
\end{enumerate}

\section{Methodology}
\begin{figure}[ht]
  \centering
  \includegraphics[width=0.75\linewidth]{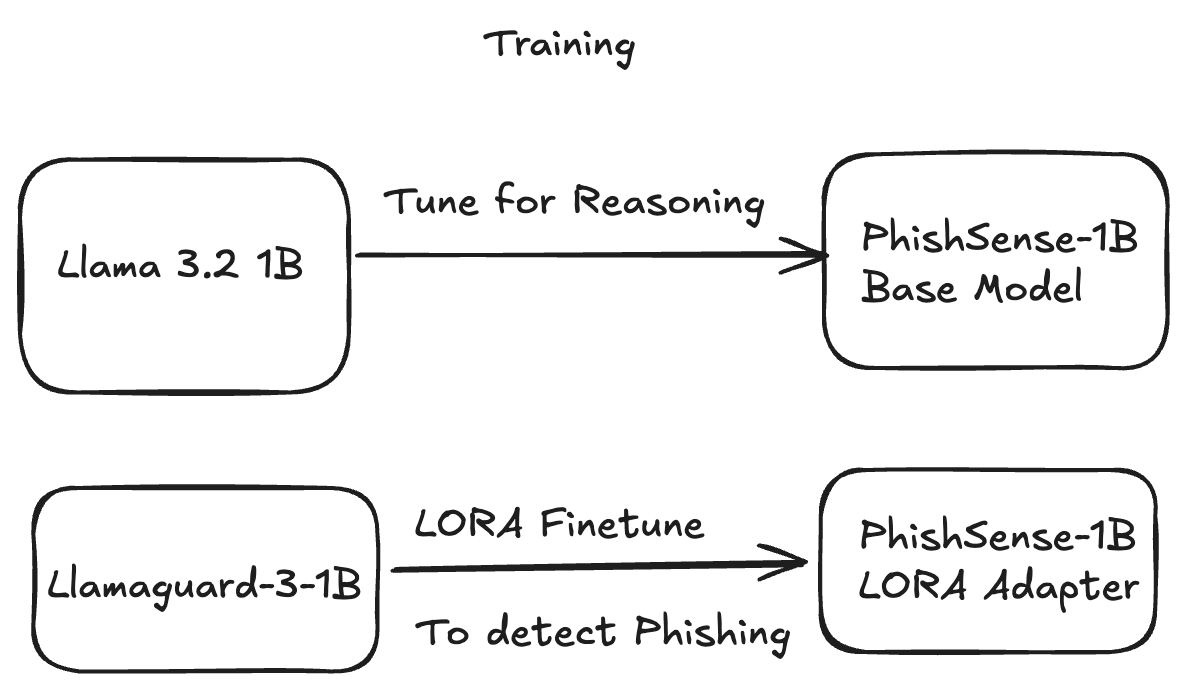}
  \caption{High-level schematic of the training workflow. The \texttt{llama-3.2-1B} model is fine-tuned for enhanced reasoning to form the PhishSense-1B base model. Separately, a LoRA-based fine-tuning on \texttt{llamaguard-3-1B} yields a phishing-focused adapter.}
  \label{fig:training_schematic}
\end{figure}

Our approach aims to build a specialized phishing detection system, referred to as \emph{PhishSense-1B}, by leveraging two key stages: (1) fine-tuning a base language model for enhanced reasoning, and (2) applying Low-Rank Adaptation (LoRA) to further adapt the model for phishing detection. Figure~\ref{fig:training_schematic} illustrates the overall training workflow, while Figure~\ref{fig:inference_schematic} shows how the final inference pipeline operates.

\subsection{Base Model Training}
We start with a pre-trained \texttt{llama-3.2-1B} model, which we adapt for improved reasoning capabilities. This initial fine-tuning step is guided by techniques similar to those proposed by GuardReasoner\footnote{\url{https://github.com/yueliu1999/GuardReasoner/?tab=readme-ov-file}}, resulting in a \emph{PhishSense-1B Base Model}. Specifically, we update the base model’s parameters on a balanced text corpus to ensure it can perform elementary reasoning tasks. This adaptation ensures that the core language model is robust enough to handle subtle linguistic nuances frequently found in phishing emails and URLs.

\subsection{LoRA-Based Phishing Adapter}
Next, we integrate a LoRA adapter to focus on phishing-specific patterns without retraining all of the base model’s parameters. We build upon the \texttt{llamaguard-3-1B} weights---an existing checkpoint known for its security-centric features---and apply LoRA to produce the \emph{PhishSense-1B LoRA Adapter}. LoRA restricts updates to low-rank matrices inserted within attention and feed-forward layers, drastically reducing the number of trainable parameters. During training, the base model’s weights are frozen, and only the LoRA adapter is updated. This approach yields a parameter-efficient solution capable of rapidly adapting to phishing scenarios.

\begin{figure}[ht]
  \centering
  \includegraphics[width=0.5\linewidth]{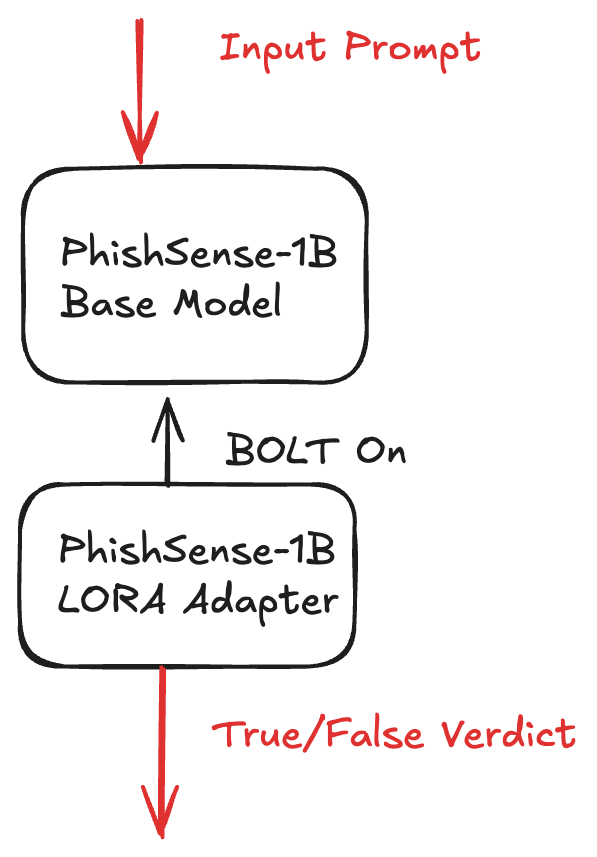}
  \caption{Inference workflow: The PhishSense-1B base model is augmented with the LoRA adapter. Only a small fraction of the parameters are updated during training, but at inference time, these components together yield a final phishing verdict.}
  \label{fig:inference_schematic}
\end{figure}

We compile a comprehensive dataset of phishing and benign samples, comprising emails, URLs, and short messages known to contain deceptive content. Stratified sampling is used to split data into training, validation, and test sets, ensuring balanced class distributions. Preprocessing includes lowercasing, removal of extraneous markup, normalization of special characters, and tokenization via a subword-based tokenizer compatible with the base model’s vocabulary. This pipeline ensures that both the base model and the LoRA adapter encounter minimal noise and maximal clarity in domain-specific patterns.

At inference time, as shown in Figure~\ref{fig:inference_schematic}, the \emph{PhishSense-1B Base Model} remains unchanged, while the LoRA adapter is “bolted on” to provide phishing-specific detection. Given an input prompt (e.g., an email body or URL text), the adapter modifies only a small subset of internal weight matrices to produce the final classification score. The system then outputs a \emph{True/False} verdict indicating whether the sample is likely phishing or legitimate.

We adopt cross-entropy loss with label smoothing and train using the AdamW optimizer at a moderate learning rate (e.g., $1 \times 10^{-3}$). Mixed-precision training further speeds convergence and reduces memory usage. By separating the “reasoning” step from the “phishing-specific” step, we ensure that the base model remains broadly competent, while the LoRA adapter rapidly assimilates phishing-related features. This design strikes a balance between model generality and domain specificity, minimizing computational overhead and overfitting risks.

\begin{figure}[ht]
  \centering
  \includegraphics[width=0.7\linewidth]{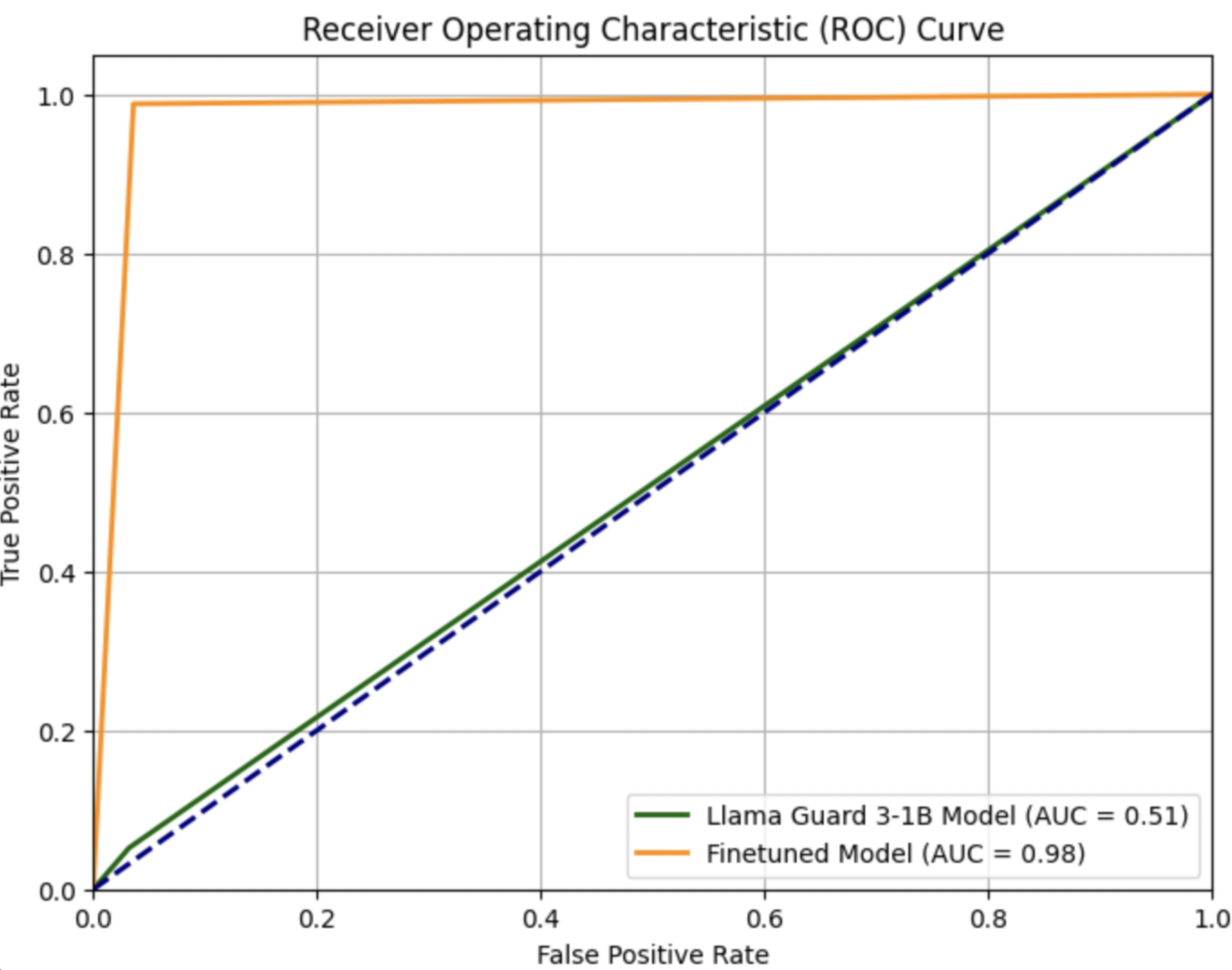}
  \caption{ROC plots comparing the base model to the adapter-inclusive model on an adversarial dataset.}
  \label{fig:llm_guard_vs_finetuned_roc}
\end{figure}

\section{Results and Discussion}
In this section, we present the quantitative performance of three models evaluated on two datasets: (1) the \emph{Custom Dataset}, (2) the \emph{RealDaten} dataset, and (3) eval generated from \emph{zefang-liu/phishing-email-dataset}. Each model’s performance is reported in terms of standard metrics, including accuracy, F1-score, precision, recall, and ROC\_AUC. Tables~\ref{tab:custom_dataset} and \ref{tab:realdaten_dataset} provide a concise overview of the numerical results.

\begin{figure}[ht]
  \centering
  \includegraphics[width=0.7\linewidth]{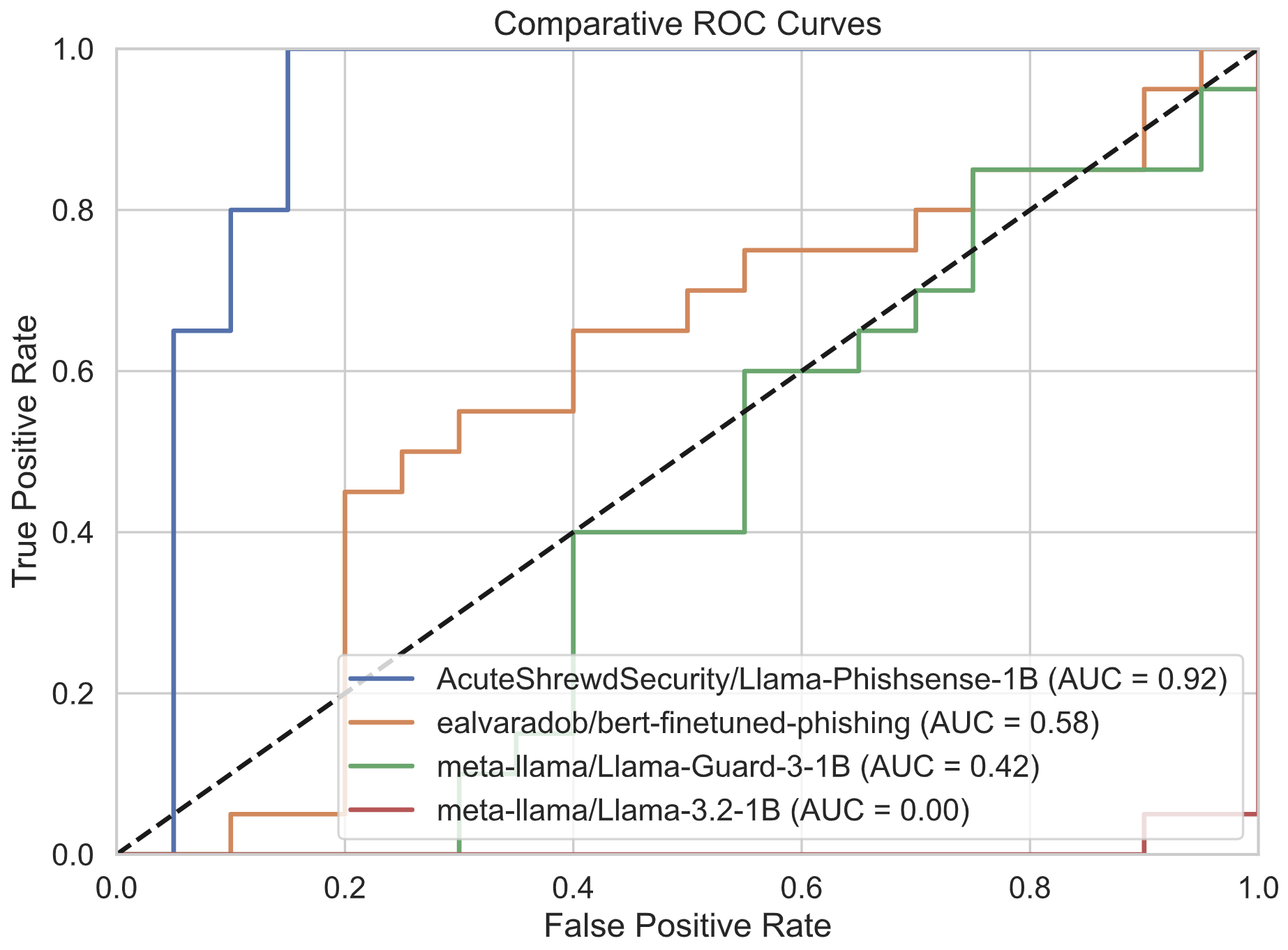}
  \caption{ROC plots comparing the base model to other comparable models in the Custom Dataset. Note that \texttt{llama-3.2-1B} is not included in other diagrams.}
  \label{fig:llamaguard_vs_finetuning}
\end{figure}

\subsection{Evaluation on efang-liu/phishing-email-dataset Dataset}
The first experiment compared \texttt{llamaguard-3-1B} and our base model by extracting a 3000 positive--3000 negative pair evaluation set from \emph{efang-liu/phishing-email-dataset}. Figure~\ref{fig:llm_guard_vs_finetuned_roc} illustrates the ROC comparison between the adapted model and the unadapted base model.

\subsection{Evaluation on Custom Dataset}
Table~\ref{tab:custom_dataset} and Figure~\ref{fig:llamaguard_vs_finetuning} show the results obtained by the three models on our custom dataset of phishing and valid emails:
\begin{itemize}
    \item \textbf{AcuteShrewdSecurity/Llama-Phishsense-1B} demonstrates the highest overall performance with an accuracy of 0.975, an F1-score of 0.976, and a ROC\_AUC of 1.0. Notably, its recall of 1.0 indicates it correctly identified all phishing samples.
    \item \textbf{ealvaradob/bert-finetuned-phishing} yields a moderate accuracy of 0.625 and an F1-score of 0.595. Its precision (0.647) is higher than its recall (0.55), suggesting that while it errs on the side of caution, it misses a nontrivial number of malicious samples.
    \item \textbf{meta-llama/Llama-Guard-3-1B} struggles significantly with an accuracy of 0.50 and an F1-score of 0.0, indicating that without adaptation, the base model fails to detect phishing samples.
\end{itemize}

\begin{table}[ht]
\centering
\caption{Model Performance on the Custom Dataset}
\label{tab:custom_dataset}
\resizebox{\linewidth}{!}{%
\begin{tabular}{lccccc}
\toprule
\textbf{Model} & \textbf{Accuracy} & \textbf{F1} & \textbf{Precision} & \textbf{Recall} & \textbf{ROC\_AUC} \\
\midrule
AcuteShrewdSecurity/Llama-Phishsense-1B & 0.975 & 0.976 & 0.952 & 1.000 & 1.000 \\
ealvaradob/bert-finetuned-phishing       & 0.625 & 0.595 & 0.647 & 0.550 & 0.585 \\
meta-llama/Llama-Guard-3-1B               & 0.500 & 0.000 & 0.000 & 0.000 & 0.000 \\
\bottomrule
\end{tabular}%
}
\end{table}

\subsubsection{Discussion for Custom Dataset}
The exceptional performance of \emph{Phishsense-1B} confirms the effectiveness of LoRA-based adaptation. By updating only a small fraction of parameters, the model achieves near-perfect recall, ensuring that no phishing email is missed---a critical requirement in cybersecurity scenarios. In contrast, the unadapted base model (\emph{Llama-Guard-3-1B}) fails entirely, underscoring the necessity of domain-specific adaptation.

\subsection{Evaluation on RealDaten Dataset}
We next assess the models on the RealDaten dataset, which contains noisier and more diverse real-world data:
\begin{itemize}
    \item \textbf{AcuteShrewdSecurity/Llama-Phishsense-1B} achieves an accuracy of 0.70 and an F1-score of 0.75, with a recall of 0.90 and a ROC\_AUC of 0.795, demonstrating robust performance even in challenging datasets.
    \item \textbf{ealvaradob/bert-finetuned-phishing} reaches an accuracy of 0.55 and an F1-score of 0.690. Although it achieves perfect recall (1.0), its precision is lower (0.526), indicating a higher false positive rate.
    \item \textbf{meta-llama/Llama-Guard-3-1B} again performs poorly with an accuracy of 0.50 and near-zero other metrics, further confirming the need for adaptation.
\end{itemize}

\begin{table}[ht]
\centering
\caption{Model Performance on the RealDaten Dataset}
\label{tab:realdaten_dataset}
\resizebox{\linewidth}{!}{%
\begin{tabular}{lccccc}
\toprule
\textbf{Model} & \textbf{Accuracy} & \textbf{F1} & \textbf{Precision} & \textbf{Recall} & \textbf{ROC\_AUC} \\
\midrule
AcuteShrewdSecurity/Llama-Phishsense-1B & 0.70 & 0.75 & 0.643 & 0.90 & 0.795 \\
ealvaradob/bert-finetuned-phishing       & 0.55 & 0.690 & 0.526 & 1.00 & 0.640 \\
meta-llama/Llama-Guard-3-1B               & 0.50 & 0.000 & 0.000 & 0.00 & 0.563 \\
\bottomrule
\end{tabular}%
}
\end{table}

\begin{figure}[ht]
  \centering
  \includegraphics[width=0.7\linewidth]{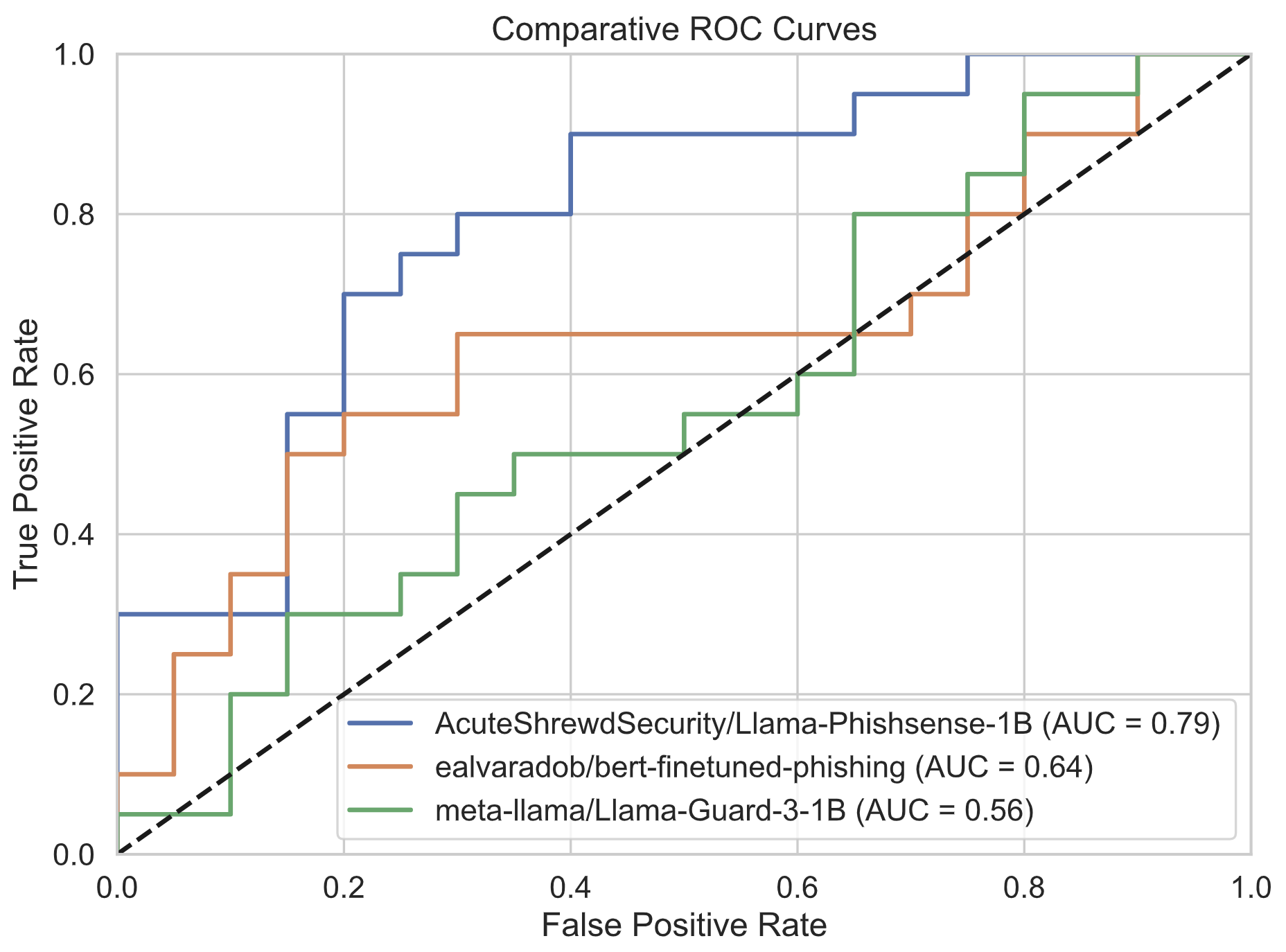}
  \caption{ROC plots comparing the base model to other comparable models in the RealDaten Dataset.}
  \label{fig:realdaten}
\end{figure}

\subsubsection{Discussion for RealDaten Dataset}
While overall accuracy is lower on the RealDaten dataset, \emph{Phishsense-1B} remains competitive, balancing high recall with acceptable precision. The BERT-based approach, despite perfect recall, suffers from an excessive false positive rate, making it less practical in environments where operational efficiency is paramount.

\section{Practical Implications and Future Work}
The consistent success of \emph{Phishsense-1B} across datasets highlights the benefits of LoRA adaptation for phishing detection. This approach enables rapid, memory-efficient fine-tuning---vital for adapting to the evolving nature of phishing attacks.

In security-sensitive environments, models with extremely high recall (like the BERT-based approach) may overwhelm security teams with false positives. In contrast, \emph{Phishsense-1B} offers a balanced solution suitable for deployment on resource-constrained devices.

Potential avenues for future research include:
\begin{itemize}
    \item \textbf{Explainable AI Techniques:} Integrating attention-based visualizations and gradient-based attribution methods to elucidate decision-making processes. We employed an avenue of creating security intelligence in this work. It is plausible that similar strategies could work in other areas such as AI-powered EDR.
    \item \textbf{Multilingual and Multimodal Extensions:} Extending the framework to non-English datasets and incorporating additional data modalities, such as visual cues.
    \item \textbf{Deployment via Browser Extensions:} Integrating the model into a Chrome extension for real-time phishing protection, thereby enhancing end-user security.
\end{itemize}

\section{Limitations}
While the results presented here are encouraging, several limitations must be acknowledged:
\begin{itemize}
    \item \textbf{Dataset Diversity:} The Custom and RealDaten datasets may not capture the full range of evolving phishing tactics, which could limit generalizability.
    \item \textbf{Evaluation Metrics:} Standard metrics (accuracy, F1, ROC\_AUC) provide an overview but may not fully reflect the operational impact of false positives and negatives.
    \item \textbf{Adversarial Robustness:} Despite the efficiency of LoRA, the model may remain vulnerable to sophisticated adversarial attacks, necessitating further robustness testing.
    \item \textbf{Static Data Limitations:} Experiments on static datasets highlight the need for dynamic, online adaptation to keep pace with rapidly evolving phishing strategies.
    \item \textbf{Interpretability:} The current framework does not fully explain individual predictions, a gap that future work should address to build trust in real-world applications.
\end{itemize}

\section{Conclusion}
In this work, we introduced \emph{Phishsense-1B}, a phishing detection model that leverages parameter-efficient Low-Rank Adaptation (LoRA) to fine-tune a large pre-trained language model. Our two-tiered approach---combining a robust base model with a lightweight, domain-specific adapter---demonstrated significant performance gains across both controlled and real-world datasets. Experimental results confirmed that \emph{Phishsense-1B} achieves high recall and balanced precision, effectively identifying phishing attempts while maintaining computational efficiency.

The practical implications of this work are substantial. Not only does it advance the state-of-the-art in phishing detection, but it also provides a clear pathway for real-world deployment via platforms such as Chrome extensions. Future work will focus on dynamic model updates, enhanced interpretability, and extending the framework to support multilingual and multimodal data, ensuring robust protection against evolving cyber threats.

The model is available on Hugging Face \cite{shrewd_security_2025}, and the evaluation data and source code will be made available on our GitHub repository.

\section{Acknowledgements}
The authors thank several model users who created eval frameworks for transformer users. They also thank the creators of \emph{ealvaradob/phishing-dataset} and \emph{ealvaradob/bert-finetuned-phishing}.

\bibliography{references}

\end{document}